# Twisted quadrupole topological photonic crystals

Xiaoxi Zhou[1,#], Zhi-Kang Lin[1,#], Weixin Lu[1], Yun Lai[2], Bo Hou[1,†], Jian-Hua Jiang[1,†]

[1]*School of Physical Science and Technology, & Collaborative Innovation Center of Suzhou Nano Science and Technology, Soochow University, 1 Shizi Street, Suzhou, 215006, China*

[2]*National Laboratory of Solid State Microstructures, School of Physics, & Collaborative Innovation Center of Advanced Microstructures, Nanjing University, Nanjing 210093, China*

[#]These authors contributed equally to this work.

[†]Correspondence and requests for materials should be addressed to <u>jianhuajiang@suda.edu.cn (Jian-Hua Jiang)</u>, <u>houbo@suda.edu.cn (Bo Hou)</u>.

## Abstract

**Topological manipulation of waves is at the heart of the cutting-edge metamaterial researches. Quadrupole topological insulators were recently discovered in two-dimensional (2D) flux-threading lattices which exhibit higher-order topological wave trapping at both the edges and corners. Photonic crystals (PhCs), lying at the boundary between continuous media and discrete lattices, however, are incompatible with the present quadrupole topological theory. Here, we unveil quadrupole topological PhCs triggered by a twisting degree-of-freedom. Using a topologically trivial PhC as the motherboard, we show that twisting induces quadrupole topological PhCs without flux-threading. The twisting-induced crystalline symmetry enriches the Wannier polarizations and lead to the anomalous quadrupole topology. Versatile edge and corner phenomena are observed by controlling the twisting angles in a lateral heterostructure of 2D PhCs. Our study paves the way toward topological twist-photonics as well as the quadrupole topology in the quasi-continuum regime for phonons and polaritons.**

# Introduction

The discovery of higher-order topology[1-26] opens a new horizon in the study of topological phenomena. Higher-order topological insulators (HOTIs)[1-26] are intriguing topological phases where topological mechanisms manifest themselves in multiple dimensions, unveiling a paradigm beyond the bulk-edge correspondence. For instance, 2D quadrupole topological insulators (QTIs)[1, 4] exhibit one-dimensional (1D) gapped edge states and zero-dimensional (0D) corner states. The concept of QTI generalizes the conventional Bloch band topology to the Wannier band topology[1, 4] which is described by the so-called "nested Wannier bands". A $\pi$-flux lattice model for QTIs was proposed in Ref. [1] which was later realized in experimental systems based on mechanical metamaterials,[10] microwave systems,[11] electric circuits,[12] and coupled optical ring resonators.[19] However, the $\pi$-flux lattice picture is incompatible with conventional subwavelength PhCs which lie at the boundary between continuous media and discrete-lattice systems. A straightforward generalization of the $\pi$-flux lattice to PhCs would fail since there is no mechanism for flux-threading. On the other hand, quadrupole topological PhCs, where light-matter interaction can be much enhanced by strong photon confinement on the edges and corners at subwavelength scales, are on demand for topological photonics in the nonlinear and quantum regimes.

Recently, twisting has been discovered as an invaluable approach toward exotic states of matter with nontrivial topology, strong correlation, or superconductivity in 2D van der Waals materials.[27-29] However, so far, there is no connection between HOTI and twisting. Moreover, the power of twisting has not been unleashed in photonics where twisting is, in fact, experimentally more accessible than in electronic systems.

In this work, we illustrate that twisting can be an efficient tool to bring about HOTI in photonics. Instead of the bilayer moiré patterns, we use a lateral heterostructure of 2D photonic crystals with opposite twisting angles to realize photonic corner states emerging from HOTIs. Interestingly, the underlying HOTI is a photonic anomalous QTI (AQTI) emerging due to the twisting-induced crystalline symmetry. Specifically, exploiting a common square-lattice PhC with trivial topology as the motherboard, we show that a particular twisting deformation can induce the AQTI phase when the mirror symmetries are removed while the glide symmetries emerge. In AQTIs, the flux-threading mechanism is not needed, and the system is not based on tight-binding models but rather rely on the quadrupole topology due to the glide symmetries in the quasi-continuum regime where the photonic bands are induced by Bragg scatterings. Moreover, the twisting approach leads to PhCs with "anomalous quadrupole topology" which differs fundamentally from the conventional quadrupole topology as follows: First, the twisting induced nonsymmorphic symmetry doubles the band representation, hence at least four bands are needed below the quadrupole topological band gap. Second, in the Wannier representation, four nondegenerate Wannier bands are required for a minimal description of the anomalous quadrupole topology. Third, due to the lack of mirror symmetry, the edge polarizations include a topological (quadrupolar) contribution and a trivial ($C_4$-symmetric) contribution.

In experiments, we use 2D subwavelength PhCs made of $Al_2O_3$ cylinders to realize the photonic AQTIs via twisting the unit-cell structure. Using the electromagnetic near-field scanning methods, we directly measure and visualize the photonic wavefunctions of the bulk, edge and corner states. We reveal that the twisting angle can effectively control the photonic corner and edge states, leading to versatile topological boundary phenomena. Exploiting such controllability, we demonstrate that when the frequencies of the edge and corner states are tuned close, their mutual couplings enable excitation of corner states via the edge states. These findings unveil the twisted quadrupole topological PhCs with rich corner and edge phenomena and may inspire future developments of twisting photonics in the up-rising field of topological photonics.[30-50]

## Results

### System and symmetry

The 2D square-lattice PhC has four identical cylinders made of $Al_2O_3$ (radius 0.25 cm) in each unit-cell. The lattice constant is $a = 2$ cm. The dielectric PhC is placed in a 2D cavity formed by metallic cladding above and below, where the transvers-magnetic harmonic modes (i.e., the modes with electric fields along the z direction) dominate the photonic bands at low frequencies. We start with a configuration where the four cylinders are located at the positions of $(\pm \frac{a}{4}, \pm \frac{a}{4})$ and the unit-cell has $C_{4v}$ point-group symmetry. By twisting the four cylinders along the dashed lines as illustrated in Fig. 1a, the crystalline symmetry is reduced to the nonsymmorphic group $P4g$ which contains the two glide symmetries, $G_x := (x, y) \to (\frac{a}{2} - x, \frac{a}{2} + y)$ and $G_y := (x, y) \to (\frac{a}{2} + x, \frac{a}{2} - y)$, the four-fold rotation symmetry, $C_4$, and the inversion symmetry $\mathcal{I} := (x, y) \to (-x, -y)$. Note that the two glide symmetries do not commute with each other, $G_x G_y \neq G_y G_x$. Here, the twist is not a pure rotation, but is designed to preserve the glide symmetries. When the dielectric cylinders overlap with each other, we combine the overlapping cylinders together.

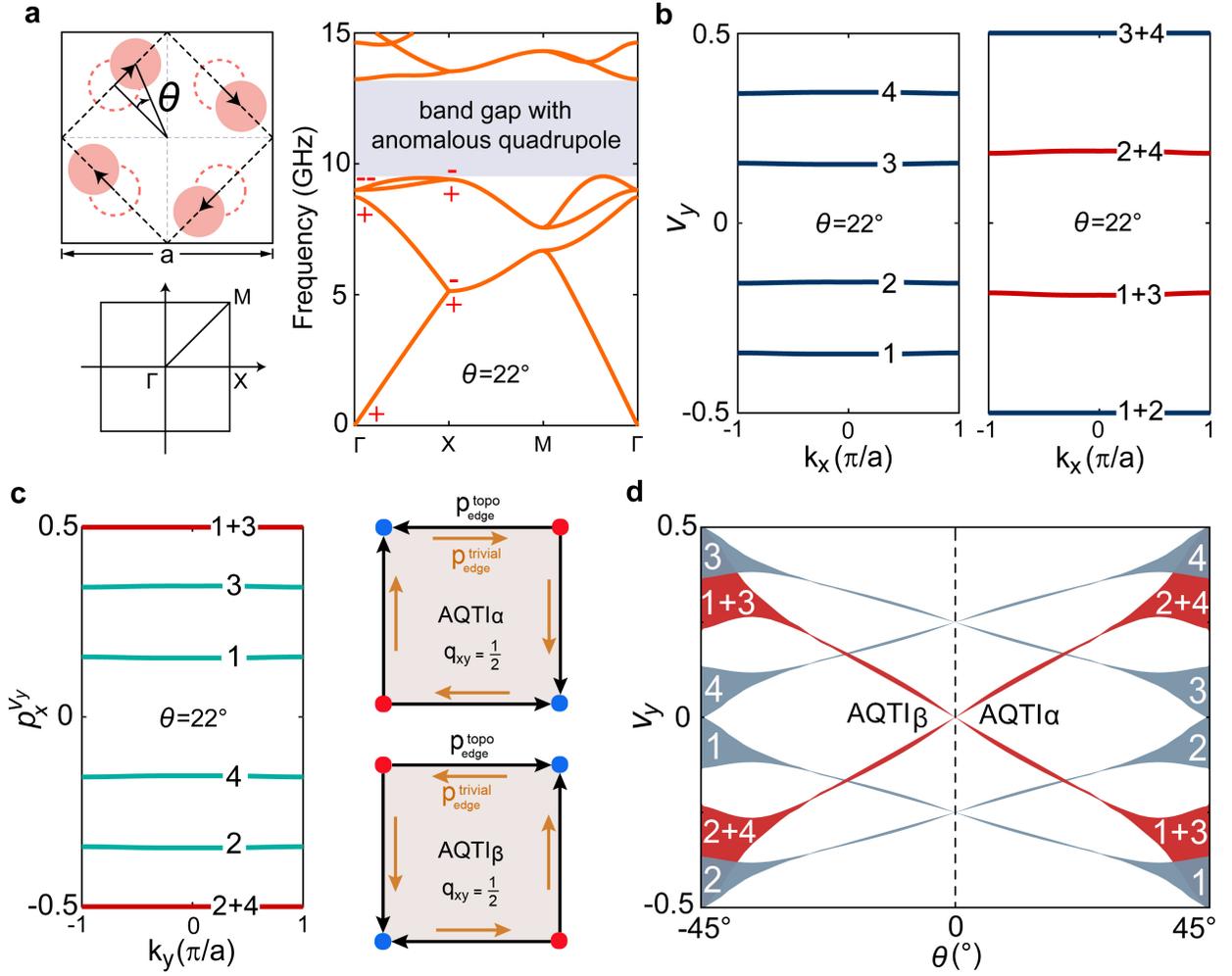

**Figure 1.** Photonic anomalous quadrupole topological insulators. (a) Left: Unit-cell structure and Brillouin zone. Twisting is along the dashed lines with $\theta$ denoting the (clockwise) twisting angle. Right: Photonic band structure for $\theta = 22°$. $\pm$ denote the parity of the photonic bands at the $\Gamma$ and X points (+/− for even/odd parity, respectively). The photonic band gap between the fourth and fifth bands exhibits the anomalous quadrupole topology. (b) Wannier bands and their combinations. (c)

Nested Wannier bands (left) and the illustration of edge polarizations and quadrupole topology (right). The edge polarizations include the topological contribution with a quadrupolar geometry and the trivial contribution with a $C_4$-symmetric geometry. (d) Evolutions of the Wannier bands and their combinations with the twisting angle $\theta$.

## Photonic bands

We now point out the consequences of the glide symmetries on the photonic bands. First, the glide symmetries lead to double degeneracy of photonic bands at the Brillouin zone boundaries (i.e., the MX and MY lines; they are equivalent since the system has the $C_4$ rotation symmetry). This can be illustrated via the anti-unitary operators $\Theta_i \equiv G_i \mathcal{T}$ ($i = x, y$). Here, $\mathcal{T}$ is the time-reversal operator which is explicitly the complex conjugation operator for the electromagnetic wavefunctions. One finds that $\Theta_x^2 \psi_{n,\vec{k}} = -\psi_{n,\vec{k}}$ if $k_x = \pi/a$, and $\Theta_y^2 \psi_{n,\vec{k}} = -\psi_{n,\vec{k}}$ if $k_y = \pi/a$, for any photonic Bloch wavefunction $\psi_{n,\vec{k}}$ with $n$ and $\vec{k}$ being the band index and the wavevector, respectively. Similar to the Kramers theorem, these algebraic properties give rise to the double degeneracy for *all* photonic bands at the MX and MY lines. In addition, the inversion operator, $\mathcal{I}$, and the anti-unitary operator, $\Theta_x$ ($\Theta_y$), anti-commutate at the X (Y) point. Therefore, the doubly degenerate bands at the X (Y) point always include an odd-parity band and an even-parity band [See Fig. 1a; See Supplementary Note 1 for the proof]. From the parity-inversion between the Γ and X (Y) points (see Fig. 1a), we conclude that the Wannier dipole is quantized to $\vec{P} = (\frac{1}{2}, \frac{1}{2})$ for the partial photonic band gap between the second and the third bands, whereas the Wannier dipole is quantized to $\vec{P} = (0,0)$ for the complete photonic band gap between the fourth and the fifth bands. The vanishing dipole polarization of the complete photonic band gap provides a necessary condition for the emergence of the quadrupole topology.

The photonic band structure for the PhC with $\theta = 22°$ is shown in Fig. 1a which exhibits a large photonic band gap (~30%) between the fourth and fifth bands, ranging from 9.46 GHz to 13.2 GHz, corresponding to a subwavelength regime (i.e., the structure features and the lattice constant are smaller than the wavelengths in free-space). Throughout this paper, the permittivity of the dielectric cylinders is taken as $\varepsilon = 6.2$ for the 2D simulation to represent the experimental measurements in the quasi-2D systems approximately (See Supplementary Note 2 for the detailed comparison between the 2D approximation and the 3D simulation).

## Quadrupole topology

We illustrate below that such a photonic band gap carries the anomalous quadrupole topology. The underlying physics of the AQTI phase does not rely on any tight-binding model, but can be directly characterized through the "nested Wannier bands" approach[1, 4] using the photonic wavefunctions from first-principle calculations (see Fig. 1). The nontrivial quadrupole topology in the $P4g$ PhCs originates from the symmetry-enforced quantization of the quadrupole polarization due to the glide symmetries (See Supplementary Note 3 for the proof).

According to Refs. [1] and [4], the quadrupole topology requests the following key elements: First, the gapped Wannier bands and the vanishing dipole polarization, meaning that the Wannier centers are away from 0 and $\frac{1}{2}$ and come in pair with both positive and negative values, i.e., $(-v, v)$ with $0 < v < \frac{1}{2}$. Second, the nontrivial, quantized quadrupole moment as manifested in the Wannier sector polarizations. The former is a necessary condition for the latter. In previous theories[1, 4], the first condition is realized via the flux-threading mechanism which leads to the non-commutative mirror symmetries. Without the flux-threading, the commutative mirror symmetries lead to gapless Wannier bands and hence forbid the quadrupole topology. In our flux-free systems, the motherboard PhC has commutative mirror symmetries and gapless Wannier bands in all combinations. When the mirror symmetries are removed by twisting, the non-commutative glide symmetries give rise to the quadrupole topology.

The Wannier bands are calculated from the photonic Bloch wavefunctions using the Wilson-loop approach. The Wilson-loop operator along, e.g., the $y$ direction is defined as [1, 4, 51]

$$\widehat{W}_{y,\boldsymbol{k}}(k_x) = \mathcal{T}_P \exp[i \oint \hat{A}^y(\boldsymbol{k}) dk_y] , \qquad (1)$$

where the subscript $y$ and $\boldsymbol{k}$ specify, respectively, the direction and the starting point of the loop. $\hat{A}^y(\boldsymbol{k})$ is the matrix (non-Abelian) formulation of the photonic Berry connection with its matrix element written as $A^y_{nm}(\boldsymbol{k}) = i \langle E_m(\boldsymbol{k}) | \partial_{k_y} | E_n(\boldsymbol{k}) \rangle$ where $|E_n(\boldsymbol{k})\rangle$ is the periodic part of the photonic Bloch wavefunction for the electric field along the $z$ direction. The ket-bra symbols and the inner product for the photonic wavefunctions are defined in Supplementary Note 4. The first four photonic bands below the topological band gap are numerated by $n, m = 1, 2, 3, 4$. $\mathcal{T}_P$ represents the path-ordering operator along a closed loop in the Brillouin zone. Here, the Wilson-loop path has fixed $k_x$, but with $k_y$ traversing the whole region of $[0, \frac{2\pi}{a}]$.

The Wannier bands are obtained by diagonalizing the Wilson-loop operator, $\widehat{W}_{y,\mathbf{k}}\xi^j_{y,\mathbf{k}} = e^{2\pi i v^j_y(k_x)}\xi^j_{y,\mathbf{k}}$ for $j = 1, 2, 3, 4$. The $j$-th Wannier band is explicitly the dependence of the Wannier center $v^j_y(k_x)$ on the wavevector $k_x$ (see Fig. 1b). The eigenvectors $\xi^j_{y,\mathbf{k}}$ are used to construct the Wannier band bases, [1] $|w_j(\mathbf{k})\rangle = \sum_{n=1}^{4}[\xi^j_{y,\mathbf{k}}]^n|E_n(\mathbf{k})\rangle$ with $[\xi^j_{y,\mathbf{k}}]^n$ denoting the $n$-th element of the eigenvector $\xi^j_{y,\mathbf{k}}$. The Wannier band bases can be regarded as the Wannier "wavefunctions" from which the topology of the Wannier bands can be defined. For instance, the polarizations of the gapped, non-degenerate Wannier bands are given by $p^{v_y}_{x,j}(k_y) = \frac{1}{2\pi}\oint \tilde{A}^x_j(\mathbf{k})dk_x$ where $\tilde{A}^x_j(\mathbf{k}) = i\langle w_j(\mathbf{k})|\partial_{k_x}|w_j(\mathbf{k})\rangle$ is the Berry connection for the $j$-th Wannier band. The Wannier band polarizations, $p^{v_y}_{x,j}(k_y)$, which characterize the topological properties of the Wannier bands, are termed as the "nested Wannier bands" (see Supplementary Note 4 for calculation details).[1]

Figure 1b shows that there are four Wannier bands which are non-degenerate and gapped, distributing symmetrically in the positive and negative regions. Unlike the conventional QTIs with two Wannier bands, the four Wannier bands in AQTIs enable rich combinations. We find that only the "1+3" and "2+4" Wannier sectors yield gapped, composite Wannier bands which are necessary for the emergence of quadrupole topology. We denote the Wannier sector "1+3" ("2+4") as I (II). Since the Wannier band is negative (positive) for the Wannier sector I (II), it is adiabatically connected with the edge states at the lower (upper) edge[1, 4]. The Wannier band polarization $P^{v_y,\text{I}}_x = P^{v_y}_{x,1} + P^{v_y}_{x,3}$ ($P^{v_y,\text{II}}_x = P^{v_y}_{x,2} + P^{v_y}_{x,4}$) then gives the topological polarization of the lower (upper) edge due to the bulk. Such analysis accounts for the topological edge polarizations of the edge band gap, which are used to describe the quadrupole polarization and the induced topological corner states[1, 4].

Remarkably, our calculation show that $P_x^{v_y,I} = -P_x^{v_y,II} = \frac{1}{2}$. Such a quantization is due to the glide symmetries, as proved in Supplementary Note 3. Since the $x$ and $y$ directions are equivalent here, one finds that $P_y^{v_x,I} = -P_y^{v_x,II} = \frac{1}{2}$. The quadrupole polarization is then quantized as $q_{xy} = 2P_x^{v_y,I}P_y^{v_x,I} = \frac{1}{2}$ (Fig. 1c). Because mirror symmetry is absent in $P4g$ crystals, the polarization of a physical edge contains both the topological contribution and the trivial contribution. Importantly, the trivial edge polarizations, due to their $C_4$-symmetric configurations, do not contribute to the corner charge or the bulk-corner correspondence[1, 4] (see Fig. 1c; see the analysis in Supplementary Note 5). We find that for the case with a negative twisting angle, the signs of the edge polarizations are flipped (see Fig. 1c for schematics and Fig. 2a for the topological edge polarization). Nevertheless, the quadrupole polarization remains nontrivial, $q_{xy} = \frac{1}{2}$. We denote the topological phase with positive $\theta$'s as AQTI$\alpha$, whereas the topological phase with negative $\theta$'s as AQTI$\beta$ (see Figs. 1c and 1d).

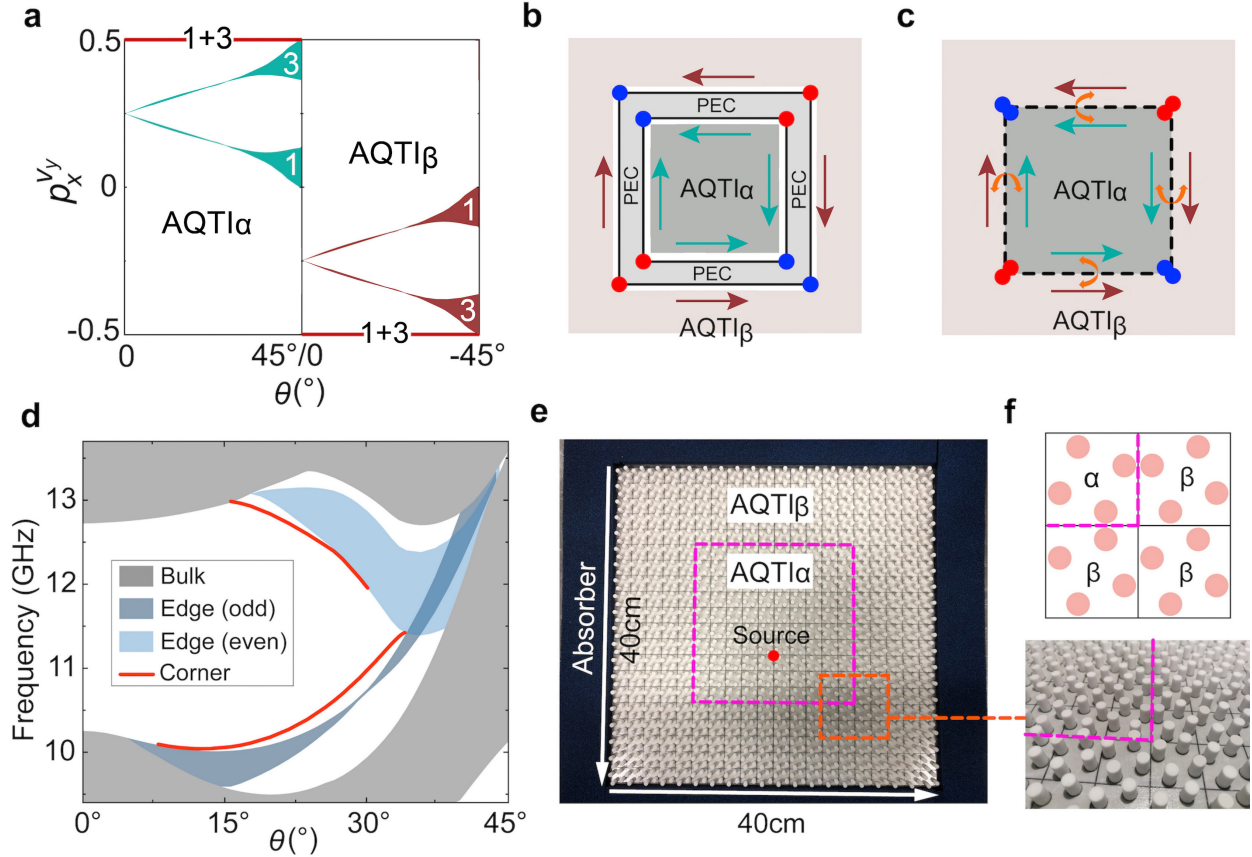

**Figure 2.** Bulk, edge and corner states modulated by the twisting angle. (a) Nested Wannier bands for AQTI$\alpha$ and AQTI$\beta$ as functions of the twisting angle. (b) Schematic illustration of the edge and corner states for a lateral heterostructure where the central region is the AQTI$\alpha$ and the outside region is the AQTI$\beta$. In between them is the region with the PEC boundaries. The arrows represent the edge states and their topological polarizations, while the red and blue dots represent the corner states. (c) The lateral heterostructure without the PEC boundaries. The boundary between the two PhCs is denoted by the black dashed lines. Orange double arrows represent interactions between the edge states from the two sides of the boundary. (d) Evolution of the bulk, edge (including odd- and even-parity edge modes) and corner states with the twisting angle $\theta$. The bulk spectrum is obtained from the unit-cell calculation, while the edge and corner spectra are from the ribbon- and box-like (as shown in (c)) supercell calculations, respectively. (e) The experimental set-up that realizes (c). (f) The detailed structure of the PhCs at the corner. Upper panel: top-down view; Lower panel: bird-view photograph.

The evolution of the Wannier bands with the twisting angle $\theta$ is shown in Fig. 1d. The Wannier sectors "1+3" and "2+4" (i.e., sectors I and II) become gapless in the limits: $\theta \to 45°$ and $\theta \to 0$ where the mirror symmetries are recovered. These properties agree with the observations in Ref. [4] that mirror-symmetric systems without flux-threading cannot support nontrivial quadrupole topology. In addition, the photonic band gap closes in the limit of $\theta \to 45°$. The twisting angle thus controls the Wannier band gap and the photonic band gap simultaneously.

Within the $P4g$ space group, we did not find any PhC with gapped Wannier bands but vanishing quadrupole polarization. Nevertheless, we can exploit the two PhCs with opposite twisting angles to construct a supercell with the edge and corner states. The scenario is illustrated in Figs. 2b and 2c. If a perfect-electric-conductor (PEC) boundary is used to separate AQTI$\alpha$ and AQTI$\beta$ as illustrated in Fig. 2b, then the topological edge (corner) states emerge at the edges (corners) for both AQTI$\alpha$ and AQTI$\beta$ (See Supplementary Note 6 for the photonic spectra and wavefunctions of the edge and corner states with PEC boundaries). Note that in Fig. 2b, the *outer* boundaries of the AQTI$\alpha$ have the same edge polarization configurations as the *inner* boundaries of the AQTI$\beta$, which is in accordance with the analysis in Fig. 1. However, the use of PEC is incompatible with our experimental system based on the transvers-magnetic harmonic modes and thus the PEC boundaries should be avoided.

The removal of the PEC boundaries leads to the coupling between the edge states from the outer boundaries of the AQTI$\alpha$ and those from the inner boundaries of the AQTI$\beta$ (see Fig. 2c). With such couplings, the hybridized edge states form the symmetric and anti-symmetric (i.e., even-parity and odd-parity) edge modes (see Supplementary Note 7 for their wavefunctions), because the edge boundaries are mirror symmetric, if AQTI$\alpha$ and AQTI$\beta$ have opposite twisting angles. In such a configuration, the evolution of the bulk, edge and corner spectra with the twisting angle $\theta$ for the set-up in Fig. 2c is presented in Fig. 2d. The corresponding experimental system is illustrated in Figs. 2e and 2f. As elaborated below, the AQTI$\alpha$ and AQTI$\beta$ form a lateral heterostructures with tunable and versatile properties of the edges and corners. For instance, the coupling between the edge states is reflected by the splitting between the symmetric and anti-symmetric modes, which can be controlled by the twisting angle $\theta$, as shown in Fig. 2d. With $\theta$ approaching 0, as the two PhCs are tuning into the same geometry, the edge states merge into the bulk (as shown in Fig. 3a and Supplementary Fig. 10). In the other limit, with $\theta$ approaching 45°, the close of the bulk band gap also leads to delocalization and annihilation of the edge states. In this process, the edge band gap closes before the bulk band gap closing [see Fig. 2d]. Later, these edge states merge into the bulk states (see Fig. 3a and Supplementary Note 8).

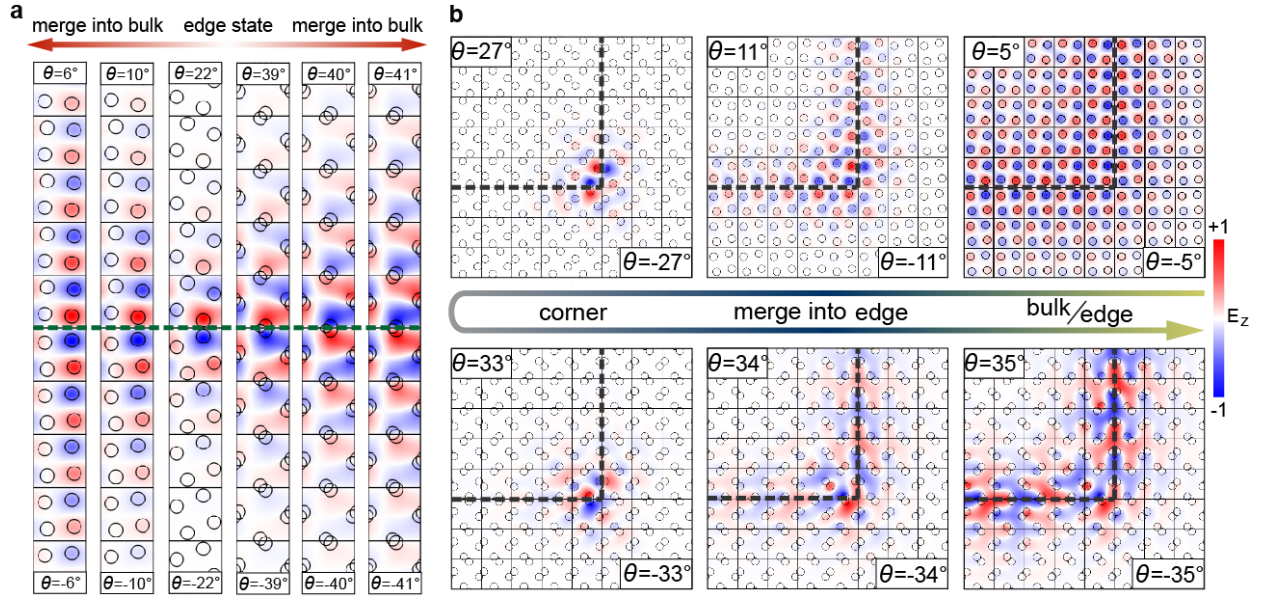

**Figure 3.** Controlling the edge and corner states by tuning the geometry. (a) Evolution of the odd-parity edge states at $k_x = 0$ with the twisting angle $\theta$. The green dashed line represents the boundary between the two PhCs with opposite twisting angles. Similar variation of the even-parity edge states is presented in the

Supplementary Figure 10. The edge states merge into the bulk in both small and large $\theta$ limit. (b) Evolution of the lower-frequency corner states with the twisting angle $\theta$. The variation of the higher-frequency corner states is presented in the Supplementary Figure 7. Note that the corner states merge into the edge in both the small and large $\theta$ limit. For the small $\theta$ limit, the corner state evolves into the odd-parity edge state, whereas in the large $\theta$ limit, it evolves into the even-parity edge state.

The corner states also experience strong modulation when the twisting angle is tuned. Fig. 3b shows the controllability of the lower-frequency corner states by the twisting angle (the study of the higher-frequency corner states is presented in the Supplementary Figure 7). With $\theta$ going from 27° to 5° the corner states gradually merge into the lower-branch (i.e., odd-parity) edge states and then into the bulk states. In the other direction, with increasing $\theta$, the corner states traverse the edge band gap and gradually merge into the upper-branch (i.e., even-parity) edge states, as $\theta \rightarrow$ 35°. Such a behavior of traversing the edge band gap is a sign of topological corner states. In contrast, the higher-frequency corner states do not have such a property. We also notice from simulations that the lower-frequency corner states are more robust than the higher-frequency corner states (see Supplementary Note 11). These findings indicate that twisting can be used as an efficient tool to transfer photonic states from 0D corner states to 1D edge states or 2D bulk states, in reconfigurable PhCs.

## Experiments

The experimental set-up used to verify the physics elaborated above is shown in Fig. 2e where the box-like PhC structure realizes the schematic illustration in Fig. 2c. The whole structure contains 20×20 unit-cells (including 10×10 unit-cells for AQTI$\alpha$ in the center and other 300 unit-cells for the AQTI$\beta$ at the outside). Electromagnetic waves are excited in the AQTI$\alpha$ region and absorbed by the absorber outside the structure. The detection for the bulk, edge and corner modes are realized by three probes located in the bulk region of AQTI$\alpha$, on the edge and at the corner, separately. They are termed as the bulk-probe, edge-probe and corner-probe, respectively.

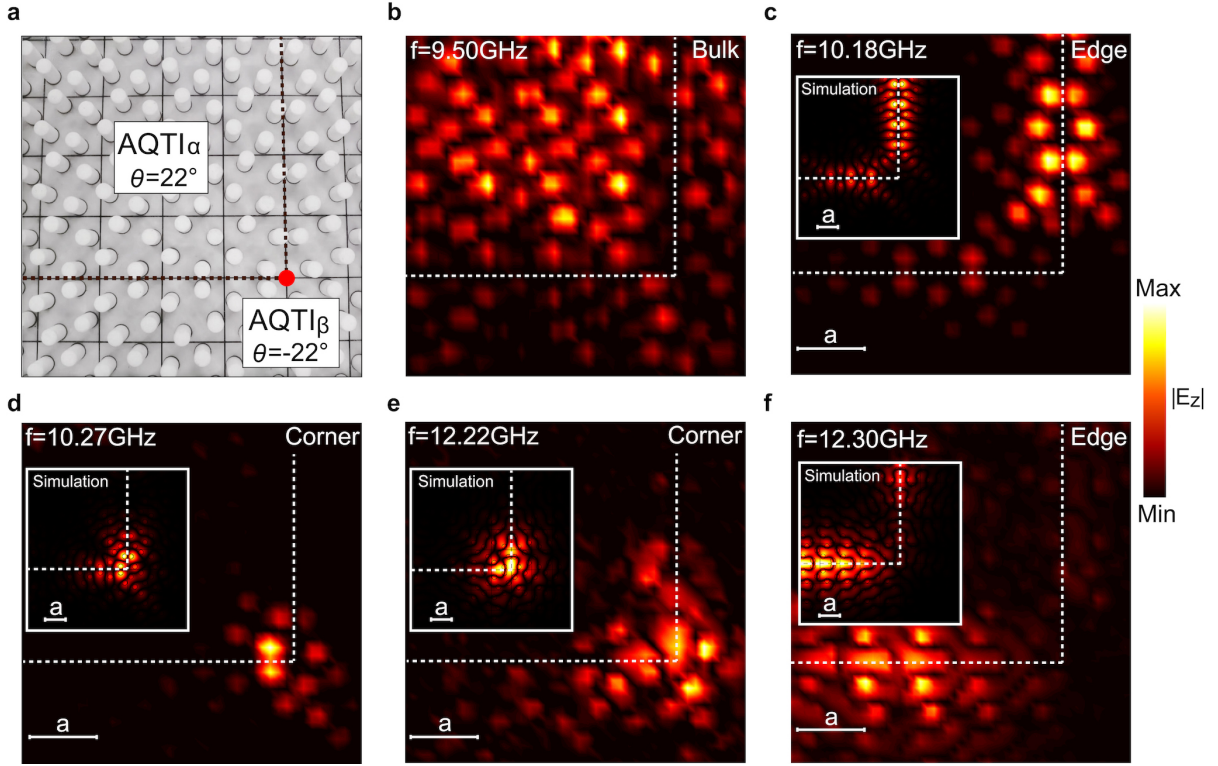

**Figure 4.** Measuring the photonic wavefunctions for the bulk, edge and corner states. (a) Photograph of a part of the lateral heterostructure used in the experiments with $|\theta| = 22°$. The corner is denoted by the red dot. (b)-(f) The measured electric field profiles for the bulk (b), edge ((c) and (f)) and corner ((d) and (e)) states. Insets: simulation results with a smaller scale. The lattice constant is labeled by the white scale bar. For all figures, the edge boundaries are denoted by the white dashed lines.

We measure the photonic wavefunction of the bulk, edge and corner states by near-field scanning of the electromagnetic field [See Supplementary Note 9]. The results are presented in Fig. 4 for a part of the lateral heterostructure with $\theta = 22°$. The photograph of the PhC structure is shown in Fig. 4a. The boundary between the two types of PhCs is illustrated by the dashed lines, while the corner is denoted by the red dot. From low frequency to high frequency, we show five typical photonic wavefunctions: the bulk state at 9.50 GHz [Fig. 3b], the edge states at 10.18 and 12.30 GHz [Figs. 3c and 3f], the corner states at 10.27 and 12.22 GHz [Figs. 3d and 3e]. The real-part of the corner wavefunctions from both the simulation and the experiments are shown in Supplementary Note 10. The robustness of the corner states against disorder is studied in Supplementary Note 11. Here, we show that the corner states In all these studies, the measured electric field distributions agree well with the numerical simulation. These results directly visualize the corner and edge states, and confirm the physics picture elaborated in the previous section.

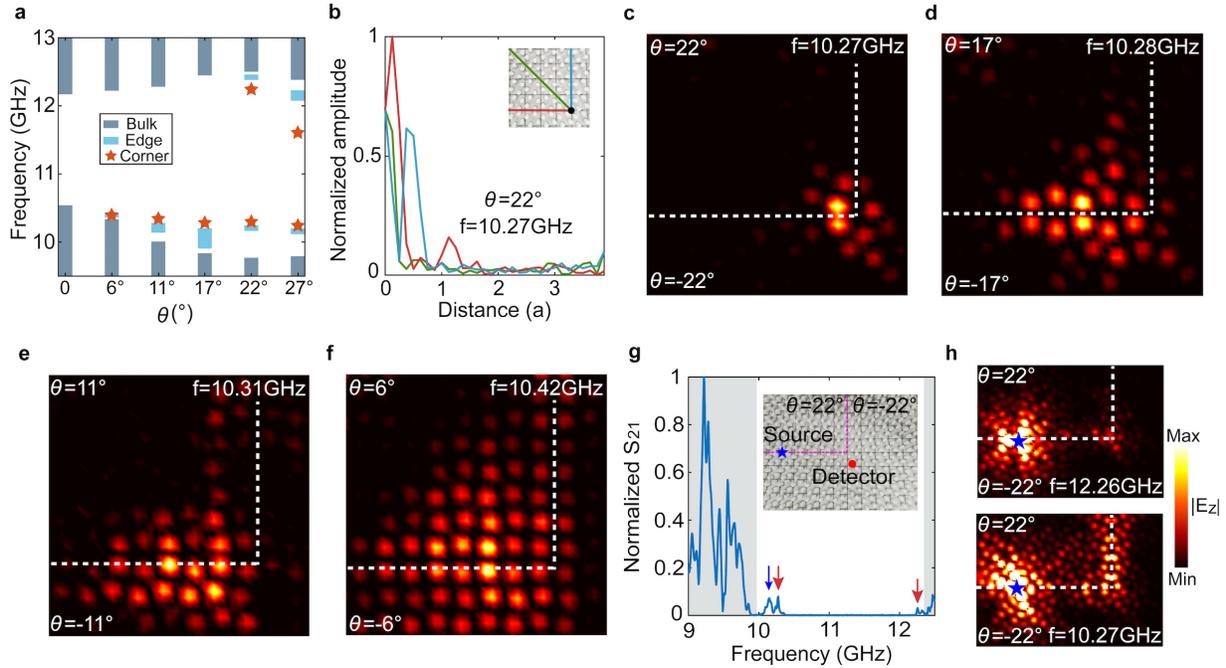

**Figure 5.** Measuring the corner states and their evolution with the twisting angle. (a) The spectral regions for the bulk, edge and corner states for various twisting angles. (b) Measurement of the electric field profiles along three directions (illustrated by the red, green and blue lines in the inset) for the corner state (resonant frequency 10.27 GHz) at $\theta = 22°$. (c)-(f) Evolution of the wavefunction of the corner states with the twisting angle. (g) Pump-probe spectroscopy for the set-up with the source at the edge and the detector close to the corner (see the inset) for $\theta = 22°$. (h) The measured electric field profile for two corner states at resonant frequencies 10.27 GHz and 12.26 GHz, respectively.

We then study the scattering coefficient ($S_{21}$) spectrum of the bulk-probe, edge-probe and corner-probe, which give effectively the local density of states in the bulk, edge and corner regions. The pump-probe spectroscopies for various twisting angles $\theta$ are presented in Supplementary Note 12. We extract from the $S_{21}$ spectra the frequency range for the bulk, edge and corner modes. The results are shown in Fig. 5a for six different twisting angles $\theta$ from 0° to 27°, which agree quite well with the simulation in Fig. 2d. Larger twisting angles are avoided because of the overlap of the $Al_2O_3$ cylinders at the unit-cell boundaries. To confirm the fully localized nature of the corner states, we measure the electric field profile of the corner state around one of the corners. Fig. 5b presents the electric field profiles measured along three lines (the red, blue and green lines as indicated in the inset) for $\theta = 22°$ at the corner mode frequency of 10.27 GHz. The results show that the photonic wavefunction is well-localized at the corner and decays rapidly in all three directions, which confirm the observation of corner states as photonic bound states.

The evolution of the lower-frequency corner state with the twisting angle $\theta$ is studied in details in Figs. 5c-5f with $\theta$ varying from 22° to 6°. These figures show clearly that the corner state gradually evolves from strongly localized to weakly localized. The corner state eventually merges into the bulk band and becomes a bulk state in the limit of $\theta \to 0$.

In the existing studies on higher-order topological insulators, the corner states are spectrally well-separated from the edge and bulk states.[1, 4, 9, 11-13, 19, 20, 25] Such isolation makes it hard to utilize them for functional devices. Here, we show that, thanks to the tunable nature of the corner states in our PhCs, the frequency of the corner states can be tuned to be close to the edge states and thus enable their mutual coupling in finite-sized systems. In experiments, we choose the set-up with the twisting angle $\theta = \pm 22°$ where the corner and edge states have frequencies, 10.27 GHz and 10.18 GHz, respectively. The coupling between the edge and corner states is revealed using the pump-probe measurement schematically illustrated in the inset of Fig. 5g. The source is placed on the edge at the left side of the corner, while the detector is placed near the corner. The pump-probe spectroscopy in Fig. 5g shows that there are two corner modes within the bulk band gap (indicated by the two red arrows): one at 10.27 GHz, the other at 12.26 GHz. The corner mode at 10.27 GHz is very close to the edge band (indicated by the blue arrow). The hybridization between the edge and corner modes is manifested directly in the measured electric field profiles in Fig. 5h which are obtained by scanning the electric field at the two frequencies, 10.27 GHz and 12.26 GHz. The electric field profile measured at 12.26 GHz indicates a clear feature of evanescent wave excitation of a spectrally-isolated, strongly-localized corner mode. In contrast, the electric field profile measured at 10.27 GHz indicates visible hybridization and coupling between the corner mode and the edge modes. Such coupled edge-corner system may serve as coupled waveguide-cavity systems in photonic chips.

# Conclusion and outlook

In this work, we unveil twisting as a new degree-of-freedom toward higher-order topology in 2D dielectric, subwavelength PhCs. Here, the nonsymmorphic symmetry induced by twisting leads to anomalous quadrupole topology for photons. The intriguing properties of the photonic anomalous quadrupole topological insulators are revealed using a lateral heterostructure comprised of two PhCs with opposite twisting angles. The photonic wavefunctions of the edge and corner states are directly visualized using the near-field scanning methods. Consistent theory and experiments show that the photonic and Wannier band gaps of the PhCs can be controlled effectively by the twisting degree-of-freedom. Consequently, rich edge and corner phenomena are observed when the twisting angle is tuned, demonstrating efficient photonic states transfer among 0D corner states, 1D edge states and 2D bulk states via twisting. With both simulation and experiments, we demonstrate that lateral heterostructures with different twisting angles can yield rich topological phenomena, and thus opens a new route toward topological photonics. Our study opens a new pathway toward twisting-photonics with higher-order topology and reconfigurable quadrupole topological photonic chips for future topological photonics.

*Note added:* At the final stage of this work, we became aware of recent works on quadrupole topological insulators in magnetized systems.[52, 53]

# Acknowledgements

Z.K.L and J.H.J are supported by the Jiangsu Province Specially-Appointed Professor Funding and the National Natural Science Foundation of China (Grant No. 11675116). Y.L is supported by National Natural Science Foundation of China (Nos. 11574226, 11704271, and 11874274). X.Z, W.L and B.H are supported by the National Natural Science Foundation of China (Grant No. 11474212).

# Competing Interests

The authors declare that they have no competing financial interests.